\documentstyle[twoside,fleqn,espcrc2,psfig]{article}
\newcommand{\beq}{\begin{equation}}
\newcommand{\eeq}{\end{equation}}
\newcommand{\beqn}{\begin{eqnarray}}
\newcommand{\eeqn}{\end{eqnarray}}
\newcommand{\de}{{\mathrm{d}}}

\newcommand{\AmS}{{\protect\the\textfont2
  A\kern-.1667em\lower.5ex\hbox{M}\kern-.125emS}}
\title{Measuring the quark condensate from the decays $\tau \rightarrow 3
\pi + \nu_{\tau}$\thanks{Talk presented at the QCD99 Euroconference,
Monpellier 7-13 July 1999. This work was supported in part by the
EEC-TMR Program, Contract N. CT98-0169 (EURODA$\Phi$NE).}}

\author{L. Girlanda\address{
Groupe de Physique Th\'eorique, Institut de Physique Nucl\'eaire, 
91406 Orsay Cedex, France.}}
\begin{document}

\begin{abstract}
The possibility of detecting the S-wave of the decays $\tau \rightarrow 3 \pi +
\nu_{\tau}$ in the threshold region is explored, with emphasis on the
sensitivity to the size of the quark antiquark condensate $\langle \bar q q
\rangle$.
\end{abstract}

\maketitle

\section{INTRODUCTION}
Despite a great deal of efforts on both the theoretical and phenomenological
side (see Ref.~\cite{efforts} and references therein), the mechanism of the
spontaneous breakdown of chiral symmetry (SB$\chi$S) in
QCD remains still unknown, in particular for what concerns the size of the
simplest order parameter of this breakdown, the quark-antiquark condensate
$\langle \bar q q \rangle$.
If most of (quenched) lattice simulations support the standard hypothesis
that SB$\chi$S is driven by the formation of a large $\langle \bar q q
\rangle $ condensate, there are indications that dynamical massless
flavors could strongly modify the situation \cite{zweig}. On one hand some unquenched
simulations seem to indicate that chiral phase transitions could happen for a
number of massless flavors $N_f$ as small as $N_f=4$ \cite{reseau}. On the
other hand, a recent sum-rule evaluation of the Zweig rule violating low
energy constant $L_6$, suggests that the quark condensate in the SU(3)
chiral limit could be considerably smaller than the same quantity in the SU(2)
chiral limit \cite{bachir}.
On the experimental side, on-going  (E865 at BNL, DIRAC at CERN) and
forthcoming (KLOE at Da$\phi$ne) facilities, will allow to pin down $\langle
\bar q q \rangle$ in the SU(2) chiral limit from low energy $\pi\pi$
observables \cite{hand}. 
Due to the enormous $\tau$-data sample already accumulated ($\sim 10^7$
$\tau$ pairs at CLEO) and to the improvements expected in the near future
\cite{perl}
(CLEOIII-CESR, BaBar,\ldots ), it is natural to explore the possibility of
measuring  $\langle \bar q q \rangle$ from $\tau$ decays, which would
provide an  independent cross-check of determinations from $\pi\pi$
scattering.\\
The sensitivity to $\langle \bar q q \rangle$ in the exclusive decays $\tau
\rightarrow 3 \pi + \nu_{\tau}$ is contained in the S-wave. The latter,
being proportional to the 
light quark mass $\hat m = (m_u + m_d)/2$, is small compared to the P-wave
contribution, which, moreover, is enhanced by the resonance $a_1$.
However it turns out that, close to threshold, the P-wave is kinematically
suppressed relatively to the S-wave, thus allowing the latter to appear
through sizeable azimuthal left-right asymmetries.
\section{KINEMATICS}
There are two different charge modes in the decay $\tau \rightarrow 3 \pi +
\nu_{\tau}$, the $2 \pi^0 \pi^-$ mode and the all charged mode, $2 \pi^-
\pi^+$, whose hadronic matrix elements, $H^{00-}_{\mu}(p_1,p_2,p_3)$ and
$H^{--+}_{\mu}(p_1,p_2,p_3)$, are
\begin{equation}
\begin{array}{l}
H^{00-}_{\mu}= \langle
\pi^{0}(p_1)\pi^{0}(p_2)\pi^{-}(p_3)|A_{\mu}^-|0\rangle, \\
H^{--+}_{\mu}= \langle
\pi^{-}(p_1)\pi^{-}(p_2)\pi^{+}(p_3)|A_{\mu}^-|0\rangle, \\
\end{array}
\end{equation}
with $A^-_{\mu}=\bar u \gamma_{\mu}\gamma_5 d$.
In the isospin limit, $m_u = m_d = \hat m$, the most general Lorentz
decomposition of each matrix element involves three form factors,
\begin{eqnarray}
&&H_{\mu}^{\mathrm{hfs}} = V_{1\mu}F_1^{\mathrm{hfs}}(p_1,p_2,p_3)  \nonumber \\
&& + V_{2\mu} F_2^{\mathrm{hfs}}(p_1,p_2,p_3)
+ V_{4\mu}
F_4^{\mathrm{hfs}}(p_1,p_2,p_3),
\end{eqnarray}
where hfs (hadronic final state) stands for $00-$ or $--+$ and
\begin{equation}
\begin{array}{l}
V_1^{\mu}=p_1^{\mu} - p_3^{\mu} -  \frac{Q(p_1-p_3)}{Q^2} Q^{\mu} ,\\
V_2^{\mu}=p_2^{\mu} - p_3^{\mu} -  \frac{Q(p_2-p_3)}{Q^2} Q^{\mu} , \\
V_4^{\mu}=p_1^{\mu} + p_2^{\mu} + p_3^{\mu} = Q^{\mu}.
\end{array}
\end{equation}
For both the hadronic final states, Bose symmetry requires that
$F_2(p_1,p_2,p_3) = F_1(p_2,p_1,p_3)$.
The form factors $F_1$ ($F_2$) and $F_4$ correspond respectively to $J=1$
and $J=0$, where $J$ is the  total angular momentum of the hadronic system
in the hadronic rest frame.
The differential decay rate is given by 
\beq \label{eq:decayrate}
\de \Gamma (\tau\rightarrow 3 \pi \nu_\tau) = \frac{(2\pi)^4}{2 M_{\tau}}
|{\mathcal M} |^2 \de \Phi_4,
\eeq
where ${\mathcal M}$ is the matrix element of the electroweak interaction
and $\de\Phi_4$ is the invariant phase space of four particles.
As shown explicitly by Kuhn and Mirkes \cite{km}, the matrix
element 
squared can be expressed in terms of 9 independent leptonic and hadronic
structure functions $L_X$ and $W_X$.
The hadronic structure functions $W_X$ only depend on the hadronic
variables, $Q^2$ and the Dalitz plot variables $s_1=(p_2 + p_3)^2$ and $s_2
= (p_1 + p_3)^2$, while all the angular dependence is relegated in the
leptonic structure functions $L_X$.
Four of the hadronic structure functions correspond to the square of the
spin~1 part of the hadronic matrix element,
\beqn
W_A &=& (x_1^2 + x_3^2) |F_1|^2 + (x_2^2 + x_3^2) |F_2|^2\nonumber \\
&& + 2 (x_1 x_2 - x_3^2) {\mathrm{Re}} (F_1 F_2^{*}), \nonumber \\
W_C &=& (x_1^2 - x_3^2) |F_1|^2 + (x_2^2 - x_3^2) |F_2|^2 \nonumber \\
 && + 2 (x_1 x_2 + x_3^2) {\mathrm{Re}}  (F_1 F_2^{*}), \nonumber \\
W_D &=& 2\left[ x_1 x_3 | F_1|^2 - x_2 x_3 |F_2|^2 \right. \nonumber \\
&& \left. + x_3 ( x_2 - x_1) {\mathrm{Re}}  (F_1 F_2^{*}) \right], \nonumber \\
W_E &=& - 2 x_3 (x_1 + x_2) {\mathrm{Im}}  (F_1 F_2^{*}),
\eeqn
one is the square of the spin~0 component
\beq
W_{SA} = Q^2 |F_4|^2,
\eeq
and the remaining ones are the interference between the spin~0 and spin~1
components
\beq
\begin{array}{l}
W_{SB} = 2 \sqrt{Q^2}  x_1 {\mathrm{Re}} (F_1 F_4^{*}) + (p_1
\leftrightarrow p_2),
\\
W_{SC} = - 2 \sqrt{Q^2}  x_1 {\mathrm{Im}} (F_1 F_4^{*}) +( p_1
\leftrightarrow p_2),
\\
W_{SD} = 2 \sqrt{Q^2} x_3  {\mathrm{Re}} (F_1 F_4^*) - ( p_1 \leftrightarrow
p_2),
\\
W_{SE} = -2 \sqrt{Q^2} x_3  {\mathrm{Im}} (F_1 F_4^*) -  (p_1
\leftrightarrow p_2),
\\
\end{array}
\eeq
The $x_i$ are kinematical functions which are linear in the pion
three-momenta in the hadronic rest frame\footnote{
This system is oriented in such a way that the $x$-axis is along the
direction of $\vec p_3$ and all the pions fly in the $x$--$y$ plane.}, $x_1 =
V_1^x$, $x_2 = V_2^x$ and
$x_3=V_1^y$: they vanish at the production threshold. Their explicit
expressions in terms of Lorentz invariant quantities can be found in
Ref.~\cite{km}.
As it is clear from the explicit expressions of the $W_X$, the purely spin~1
structure functions are suppressed, at threshold, by two powers of the
$x_i$'s, the interferences of spin~1 and spin~0 by only one power whereas the
purely spin~0 structure function $W_{SA}$ is not suppressed at all.
This is the reason for expecting a large sensitivity to the S-wave, and then
to $\langle \bar q q \rangle$ in the threshold region.
\section{GENERALIZED CHIRAL PERTURBATION THEORY}
\begin{figure} 
\centerline{\psfig{figure=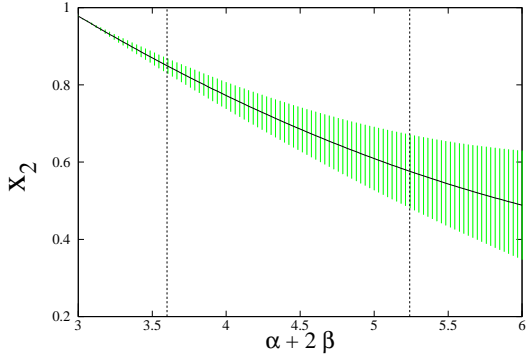,height=5.0cm,angle=-90}}
\caption{\label{fig:xvsab} The Gell-Mann--Oakes--Renner ratio as function of
$\alpha + 2 \beta$.}  
\end{figure}

The appropriate tool for studying the dependence of the form factors on
$\langle \bar q q \rangle$ at low energy is the generalized version of
chiral perturbation theory (G$\chi$PT) \cite{gchpt}.
The results of the explicit calculation of the form factors at one-loop level
can be found in Ref.~\cite{tau3pi}. Taken in the standard $\chi$PT limit
\cite{schpt}, they reproduce the results of
Ref.~\cite{colatau}.
It is convenient to parametrize the dependence on $\langle \bar q q \rangle$
of the form factors of $\tau \rightarrow 3 \pi+\nu_{\tau}$ 
in terms of two parameters $\alpha$ and $\beta$, related to the elastic
$\pi\pi$ scattering amplitude.
At leading order\footnote{
For the definition of $\alpha$ and $\beta$ up to two-loop level see
Ref.~\cite{pipi1}.}
they correspond respectively
to the amplitude and the slope at the symmetrical point $s=t=u=4/3 M_{\pi}^2$, 
\beq \label{eq:pipiampl}
A(s|t,u) = \alpha\frac{M_{\pi}^2}{3 F_{\pi}^2} + \frac{\beta}{F_{\pi}^2}
\left( s - \frac{4}{3} M_{\pi}^2 \right) +\ldots .
\eeq
In fact $\alpha$ and $\beta$ are 
correlated quantities: they must satisfy the so-called Morgan-Shaw universal
curve, which results from the analysis of Roy equations (see
Ref.~\cite{pipi1} and references therein).
All the dependence on $\langle \bar q q \rangle$ is contained in the
parameter $\alpha$: at leading order, it varies between 1 and 4 if the
quark condensate decreases from its standard value down to zero, while
$\beta$ stays always close to 1. 
The relationship between the quark condensate in the SU(2) chiral limit and
$\alpha$ and $\beta$ at one-loop level, with the associated theoretical
uncertainties, is 
shown in Fig.~\ref{fig:xvsab}, where we plot the Gell-Mann--Oakes--Renner
ratio,
\beq
x_2 =- \frac{2 \hat m}{F_{\pi}^2 M_{\pi}^2} \lim_{\hat m \to 0, m_s \neq 0}
\langle \bar q q \rangle,
\eeq
as function of $\alpha + 2 \beta$.
Also shown in the figure, between the dashed lines, is the experimentally
allowed range  for the combination $\alpha + 2 \beta$, as extracted from a
fit of the two-loop $\pi\pi$ amplitude to the $K_{e4}$ Rosselet data
\cite{rosselet}. The latter fit yields for 
$\alpha$ and $\beta$ the result \cite{pipi1},
\beq \label{eq:alfabetaexp}
\alpha^{\mathrm{exp}} = 2.16 \pm 0.86, \hspace{5mm}
\beta^{\mathrm{exp}}=1.074 \pm 0.050,
\eeq
while the predictions of standard $\chi$PT read \cite{l1l2}
\beq \label{eq:alfabetast}
\alpha^{\mathrm{st}} =1.07 \pm0.01,\hspace{5mm}
\beta^{\mathrm{st}}=1.105\pm0.015.
\eeq
\section{AZIMUTHAL ASYMMETRIES}
\begin{figure} 
\centerline{\psfig{figure=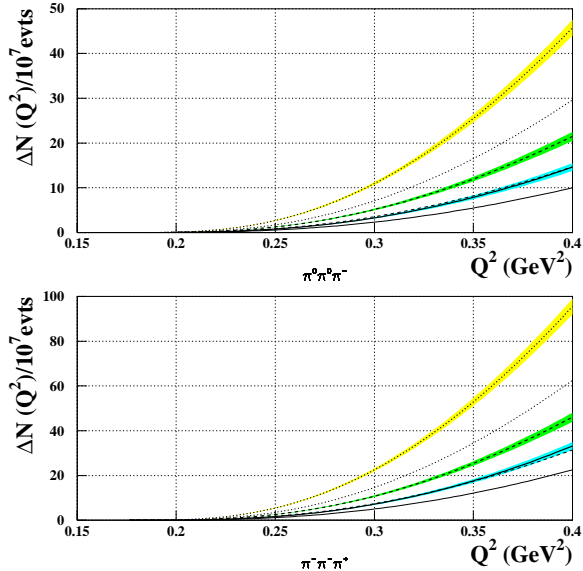,height=7.5cm,angle=0}}
\caption{\label{fig:azi} The integrated left right azimuthal asymmetry for
the two charge modes.}  
\end{figure}

The hadronic structure functions which we are interested in are the
interferences between the spin~0 and spin~1 amplitudes.
These structure functions contain the dependence on $\langle \bar q q
\rangle$ through the S-wave component and, at the same time, they are not as
small as the purely spin~0 structure function, which is suppressed by two
powers of the light quark mass $\hat m$.
If the $\tau$ rest frame can not be reconstructed (which is always the case
except for  $\tau$-charm factories), only two of them are measurable,
$W_{SB}$ and $W_{SD}$. They are related to the azimuthal asymmetries,
obtained by integrating the decay rate over all angles except the azimuthal
angle $\gamma$ (for the definitions of all angular variables see
Ref.~\cite{km}),
\beqn
\de \Gamma &=& \frac{G_{\mathrm{F}}^2}{128 M_{\tau}} \frac{V^2_{\mathrm{ud}}}{(2 \pi)^5}
 \left[ \frac{ M_{\tau}^2 - Q^2}{Q^2} \right]^2
\frac{ M_{\tau}^2 + 2 Q^2}{3 M_{\tau}^2}\, \nonumber \\
&& \times f(Q^2,\gamma)\,W(Q^2) \de Q^2
\frac{\de \gamma}{2 \pi},
\eeqn
with $W(Q^2)$  defined by
\beq
W(Q^2) = w_A + \frac{3 M_{\tau}^2}{M_{\tau}^2 + 2 Q^2} w_{SA},
\eeq
and the azimuthal distribution, normalized to~1, 
\beqn \label{eq:azdistr}
f(Q^2,\gamma) &=& 1 + \lambda_2  \left(
C^{\prime}_{\mathrm{LR}}\,\cos 2 \gamma + C^{\prime}_{\mathrm{UD}}\, \sin 2 \gamma
\right) \nonumber \\
&& + \lambda_1  \left( C_{\mathrm{LR}} \, \cos \gamma
+ C_{\mathrm{UD}}
\, \sin \gamma \right).
\eeqn
We have denoted by lowercase letters $w_X$ the corresponding structure
functions $W_X$ integrated over the whole Dalitz plot.
The asymmetry coefficients $C^{\prime}_{\mathrm{LR}}$,
$C^{\prime}_{\mathrm{UD}}$, $C_{\mathrm{LR}}$ and $C_{\mathrm{UD}}$
in Eq.~(\ref{eq:azdistr}) are related to the Kuhn and Mirkes' structure
functions by the relations\footnote{
We are neglecting the  polarization of the $\tau$'s, which is  justified
 provided that the $Z$ exchange can be neglected  (far from the $Z$ peak).} 
\beq
\begin{array}{l}
C^{\prime}_{\mathrm{LR}} = \frac{1}{3} \left( 1 - \frac{Q^2}{M_{\tau}^2}
\right) \frac{3 M_{\tau}^2}{M_{\tau}^2 + 2 Q^2} \frac{w_C}{W}, \nonumber \\
 C^{\prime}_{\mathrm{UD}} = - \frac{1}{3} \left( 1 - \frac{Q^2}{M_{\tau}^2}
 \right)  \frac{3 M_{\tau}^2}{M_{\tau}^2 + 2 Q^2}  \frac{w_D}{W}, \nonumber 
 \\
 C_{\mathrm{LR}} = - \frac{\pi}{4}  \frac{3 M_{\tau}^2}{M_{\tau}^2 + 2
Q^2}  \frac{w_{SB}}{W}, \nonumber \\
C_{\mathrm{UD}} =
\frac{\pi}{4}  \frac{3 M_{\tau}^2}{M_{\tau}^2 + 2 Q^2}  \frac{w_{SD}}{W}.
\end{array}
\eeq
The coefficients $\lambda_1$ and $\lambda_2$ \cite{marbella} are kinematical
functions 
resulting from the integration over the $\tau$-decay angle. They
depend on $Q^2$ and on the $\tau$ velocity. In the limit of ultrarelativistic
$\tau$ (relevant {\em e.g.} for CLEO), they take the form
\beqn \label{eq:lambdaas}
\lambda_1(Q^2) &=& \frac{
M_{\tau}^4 - Q^4 + 2 M_{\tau}^2 Q^2 \log \frac{Q^2}{M_{\tau}^2}
}{\left( M_{\tau}^2 - Q^2 \right)^2} , 
\nonumber \\
\lambda_2(Q^2) &=& -2 + 3 \frac{M_{\tau}^2 + Q^2}{M_{\tau}^2 - Q^2}
\lambda_1 (Q^2).
\eeqn
The explicit calculation of the form factors at one-loop level shows that
the structure function $w_{SD}$ vanishes at leading order, unlike $w_{SB}$.
The latter governs the left-right asymmetry. 
In Fig.~\ref{fig:azi} we plot the quantity $\Delta N(Q^2)= |N_L(Q^2) - N_R
(Q^2)|$ for a total number of $10^7$ $\tau$-pairs produced, where
$N (Q^2)$  denotes the total number of events, integrated from
threshold up to $Q^2$, and the subscript $L$ ($R$) refers to events with
$\pi/2 \leq \gamma \leq 3 \pi/2$  (or the complementary interval).
In order to study the sensitivity to $\langle \bar q q \rangle$ this
quantity is plotted for three values of $\alpha$ and $\beta$: for each
charge mode the lower
band corresponds to the standard predictions~(\ref{eq:alfabetast}), the middle
band to the central ``experimental''
values~(\ref{eq:alfabetaexp}) and the upper band to $\alpha =4$,
$\beta=1.16$.
For both the charge modes $\Delta N (0.35 {\mathrm{ GeV}}^2)$ changes by  a
factors~3 if $\alpha$ varies between its standard value and~4.
Quantitatively the effect is larger for the all charged mode, large enough
to be eventually detected with a sample of $10^7$ $\tau$-pairs.
Each band represents the theoretical uncertainties coming from the low energy
constants, but not from higher order chiral corrections.
The influence of the latters is estimated by studying the convergence of the
chiral series. The lines immediately below the bands correspond to the
result at tree level. We see that the one-loop corrections amount
approximately to 30\% at $Q^2=0.4$~GeV$^2$. Therefore, supposing a
geometrical behavior of the $\chi$PT series, we expect the two-loop
corrections to modify the results by no more that 10\%.
\section{CONCLUSIONS}
We have identified the decays $\tau \rightarrow 3 \pi+\nu_{\tau}$  in the
threshold region as a
promising observable for measuring the strength of the quark condensation in
QCD. A total statistics of $10^7$ $\tau$-pairs seems sufficient to
provide an interesting cross-check of future determinations of $\langle \bar
q q \rangle$ from $\pi\pi$ observables.

\end{document}